\documentclass{eas}
\usepackage{graphicx}
\begin{document}

%%-----------------------------
%%      the top matter
%%-----------------------------
\title{Advances in secular magnetohydrodynamics\\ of stellar interiors dedicated\\ to asteroseismic spatial missions} 
\author{S. Mathis$^{1,}$}
\address{CEA/DSM/DAPNIA/Service d'Astrophysique, CEA/Saclay; AIM-Unit\'e Mixte de Recherche CEA-CNRS-Universit\'e Paris VII, UMR 7158; F-91191 Gif-sur-Yvette Cedex, France}
\address{LUTH, Observatoire de Paris; 5 place Jules Janssen, F-92 195 Meudon Cedex, France}
\author{P. Eggenberger$^{3,}$}
\address{Observatoire de Gen\`eve; 51 chemin des Maillettes, CH-1290 Sauverny, Switzerland}
\address{Institut d'Astrophysique et de G\'eophysique, Universit\'e de Li\`ege; All\'ee du 6 Ao\^ut, 17 1/14 (B5C), B-4000 Li\`ege 1 (Sart Tilman), Belgium}
\author{T. Decressin$^{3}$}
\author{A. Palacios}
\address{GRAAL, Universit\'e de Montpellier II; Place Eug\`ene Bataillon, F-34095 Montpellier Cedex 05, France}
\author{L. Siess}
\address{Institut d'Astronomie et d'Astrophysique, Universit\'e Libre de Bruxelles; Campus de la Plaine, Boulevard du Triomphe, CP 226, B-1050 Bruxelles, Belgium}
\author{C. Charbonnel$^{3,}$}
\address{LATT, Observatoire Midi-Pyr\'en\'ees; 14 Avenue Edouard Belin, F-31 400 Toulouse, France}
\author{S. Turck-Chi\`eze$^{1}$}
\author{J.-P. Zahn$^{2}$}

\begin{abstract}
With the first light of COROT, the preparation of KEPLER and the future helioseismology spatial projects such as GOLF-NG, a coherent picture of the evolution of rotating stars from their birth to their death is needed. We describe here the modelling of the macroscopic transport of angular momentum and matter in stellar interiors that we have undertaken to reach this goal. First, we recall in detail the dynamical processes that are driving these mechanisms in rotating stars and the theoretical advances we have achieved. Then, we present our new results of numerical simulations which allow us to follow in 2D the secular hydrodynamics of rotating stars, assuming that anisotropic turbulence enforces a shellular rotation law. Finally, we show how this work is leading to a dynamical vision of the Hertzsprung-Russel diagram with the support of asteroseismology and helioseismology, seismic observables giving constraints on the modelling of the internal transport and mixing processes. In conclusion, we present the different processes that should be studied in the next future to improve our description of stellar radiation zones.
\end{abstract}

\runningtitle{Advances in secular magnetohydrodynamics of stellar interiors}

\maketitle

\section{Dynamics of stellar radiation zones and differential rotation}

Rotation, and more precisely the differential rotation, has a major impact on the internal dynamics of stars.\\ First, as it is known from the theory of rotating stars, rotation induces some-large scale circulations, both in radiation and convection zones, which act to transport simultaneously the angular momentum, the chemicals but also the magnetic field by advection. In radiation zones, the large-scale circulation, which is called the meridional circulation, is due to the differential rotation, to the transport of angular momentum and to the action of the perturbing forces, namely the centrifugal force and the Lorentz force (cf. Busse 1982, Zahn 1992, Maeder \& Zahn 1998, Garaud 2002, Rieutord 2006). Next, the differential rotation induces hydrodynamical turbulence in radiative regions through various instabilities: the secular and the dynamical shear instabilities, the baroclinic and the multidiffusive instabilities. In the same way that the atmospheric turbulence in the terrestrial atmosphere, this hydrodynamical turbulence acts to reduce the gradients of angular velocity and of chemical composition; it is why it is modelled like a diffusive process (cf. Talon \& Zahn 1997, Garaud 2001, Maeder 2003). On the other hand, rotation has a strong impact on the stellar magnetism. For example, it interacts with the turbulent convection in convective envelopes of solar-type stars (cf. Brun et al. 2004) to lead to a dynamo mechanism and, as it is expected from observations, to a cyclic magnetism. In radiation regions, it interacts with fossil magnetic fields where the secular torque of the Lorentz force and the magnetohydrodynamical instabilities such as the Tayler-Spruit instability and the multidiffusive magnetic instabilities have a strong impact on the transport of angular momentum and of chemicals (cf. Charbonneau \& Mac Gregor 1993, Garaud 2002, Spruit 1999, Spruit 2002, Menou et al. 2004, Maeder \& Meynet 2004, Eggenberger et al. 2005, Braithwaite \& Spruit 2005, Braithwaite 2006, Brun \& Zahn 2006). Finally, waves constitute the last transport process in single stars and they are also interacting with rotation. Internal waves, which are excited at the borders with convective zones, propagate inside radiation zones where they extract or deposit angular momentum where they are damped leading to a modification of the angular velocity profile and thus of the chemicals distribution (cf. Goldreich \& Nicholson 1989, Talon et al. 2002, Talon \& Charbonnel 2003-2004-2005, Rogers \& Glatzmaier 2005). Note also that rotation modifies stellar winds and mass losses (cf. Maeder 1999)\\ 
In close binary systems, where the companion could be a star as well as a planet, there are transfers of angular momentum between the star, its companion and the orbit due to the dissipation acting on flows induced by the tidal potential; that could be the equilibrium tide (cf. Zahn 1966) due to the hydrostatic adjustement of the star or the dynamical tide which is due to the tidal excitation of internal waves (cf. Zahn 1975). This dynamical evolution modifies the internal rotation of each component that have consequences on the properties of their internal transport.\\
To conclude, all the processes, with which rotation interacts, transport angular momentum and matter that modify the internal angular velocity, the chemical composition and the nucleosynthesis. Therefore, differential rotation has imperatively to be taken into account to get a coherent picture of the internal dynamics and of the evolution of the stars.

\section{Advances in secular magnetohydrodynamics of stellar radiation zones}

First, we have studied the rotational transport where the angular momentum and the chemicals are transported by the meridional circulation and by the hydrodynamical turbulence due to shear instabilities, namely the rotational transport of type I. We generalize its present modelling to treat simultaneously the bulk of radiation zones and their interfaces with convective zones, the tachoclines (cf. Mathis \& Zahn 2004). Next, we derive a new prescription for the horizontal turbulent transport which is obtained from Couette-Taylor laboratory experiments that allow to study turbulence in differentially rotating flows (cf. Mathis et al. 2004). However, the introduction of these two hydrodynamical mechanisms in stellar models fail to reproduce the observations of solar-type stars, because these have been slowed down by the wind during their evolution and hence the rotational processes are less efficient. Therefore, we consider the rotational transport of type II where the chemicals are still transported by the meridional circulation and the turbulence, but where the angular momentum is carried by another process, the two candidates being the magnetic field and the internal waves. So, we introduce the effect of a fossil magnetic field in a consistent way in taking into account the action of turbulence, differential rotation and meridional circulation on the field, but also its feed-back on momentum and heat transports (cf. Mathis \& Zahn 2005). Next, we introduce in the modelling of the internal waves the effect of the Coriolis force (cf. Mathis \& Zahn 2005) that allows to include the gravito-inertial waves in the description of the angular momentum transport. Finally, a coherent treatment of tidal processes has been derived. A detailed review of previous results obtained both on the rotational transport of type I and of type II is done by J.-P. Zahn in this volume.

\section{Modelling}

To get a coherent dynamical description of stellar radiation zones, the complete equations of magnetohydrodynamics have to be solved. To achieve this goal, assumptions are made.\\

\subsection{Main assumptions}

Here, secular magnetohydrodynamics and its consequences on stellar evolution are studied. Thus, secular time-scales associated to the nuclear evolution of stars are chosen. Moreover, low angular resolution (with an expansion in few spherical harmonics) is considered due to the turbulent transport behaviour in radiation zones (cf. next paragraph). Physical processes which have dynamical time-scales and need a high angular resolution description, such as hydro or magnetohydrodynamical instabilities and turbulence, are treated using prescriptions. This is the first step to achieve the highly multi-scales problem in time and in space of the dynamical stellar evolution that could not be yet studied with Direct Numerical Simulations.\\

On the other hand, stellar radiation zones are stably stratified regions. Thus, the buoyancy force, which is the restoring force, acts to inhibit turbulent motions in the vertical direction. This leads to a strongly anisotropic turbulent transport for which that in the horizontal direction (on an isobar) is more efficient than that in the vertical one. Therefore, horizontal eddy-transport coefficients are larger than those in the vertical direction and the horizontal gradients of scalar fields such as rotation, temperature and chemical concentration are smaller than their vertical gradients. An horizontal expansion in few spherical harmonics is thus allowed, and we get for a scalar field, X:
\begin{equation}
X\left(r,\theta,\varphi\right)=\overline{X}\left(r,t\right)+\delta X\left(r,\theta,\varphi,t\right)
\end{equation}
where $\delta X\left(r,\theta,\varphi,t\right)=\sum_{l>0}\sum_{m=-l}^{l}\widetilde{X}_{m}^{l}\left(r,t\right)Y_{l}^{m}\left(\theta,\varphi\right)$ with $\overline{X}\left(r,t\right)>\!\!>\widetilde{X}_{m}^{l}\left(r,t\right)$, $\overline{X}$ and $\delta X$ being respectively its horizontal average on an isobar and its fluctuation. $r,\theta,\varphi$ are the classical spherical coordinates while $t$ is the classical time.\\

The dynamical equations such as the induction equation for the magnetic field or the Navier-Stockes equation are three-dimensional vectorial equations. Here, the aim is to study the secular magnetohydrodynamics of rotating stars and its consequences on stellar structure and evolution using stellar evolution codes which are mostly unidimensional. To couple them with transport equations, we expand vector fields such as macroscopic velocities or magnetic field in vectorial spherical harmonics (see Rieutord 1987) like in stellar oscillations theory:
\begin{eqnarray}
\vec u\left(r,\theta,\varphi,t\right)=\sum_{l=0}^{\infty}\sum_{m=-l}^{l}\left\{u_{m}^{l}\left(r,t\right)\left[Y_{l}^{m}\left(\theta,\varphi\right)\widehat{e}_{r}\right]+v_{m}^{l}\left(r,t\right)\left[\vec\nabla_{\rm H}Y_{l}^{m}\left(\theta,\varphi\right)\right]\right.\nonumber\\
+{\left.w_{m}^{l}\left(r,t\right)\left[\vec\nabla_{\rm H}\wedge(Y_{l}^{m}\left(\theta,\varphi\right)\widehat{e}_{r})\right]\right\}}\hbox{ }\hbox{where}\hbox{ }\vec\nabla_{\rm H}=\widehat{e}_{\theta}\partial_{\theta}+\widehat{e}_{\varphi}\frac{1}{\sin\theta}\partial_{\varphi}.
\end{eqnarray}

These expansions allow to separate variables in transport equations. Thus, modal equations in $r$ and $t$ only are obtained and can be implemented directly in stellar evolution codes.

\subsection{Preliminary definitions}

The macroscopic velocity field is expanded as:
\begin{equation}
\vec V=r\sin\theta\Omega\left(r,\theta\right){\widehat e}_{\varphi}+\dot r{\widehat e}_{r}+\vec{\mathcal U}_{M}\left(r,\theta\right)+\vec u\left(r,\theta,\varphi,t\right). 
\end{equation}
The first term, where $\Omega\left(r,\theta\right)$ is the internal angular velocity and ${\widehat e}_{\varphi}$ is the unit vector in the $\varphi$ direction, is the azimuthal velocity field associated to the differential rotation. The second term corresponds to the radial lagrangian velocity due to the contractions and the dilatations of the star during its evolution, ${\widehat e}_{r}$ being the radial unit vector. The third term $\vec{\mathcal U}_{M}\left(r,\theta\right)$ is the meridional circulation velocity field which has been presented before. Following the general method concerning the expansion of vector fields, it is projected on vectorial spherical harmonics:
\begin{equation}
\vec{\mathcal U}_{M}=\sum_{l>0}\left\{U_{l}\left(r\right)P_{l}\left(\cos\theta\right){\widehat e}_{r}+V_{l}\left(r\right)\frac{{\rm d}P_{l}\left(\cos\theta\right)}{{\rm d}\theta}{\widehat e}_{\theta}\right\}.
\end{equation}
The anelastic approximation is adopted, thus filtering-out sonic waves, that is justified for the slow meridional circulation. Therefore, the continuity equation $\partial_{t}\rho+\vec\nabla\cdot\left(\rho\vec V\right)=0$, where $\rho$ is the density, becomes $\vec\nabla\cdot\left(\rho\vec{\mathcal U}_{M}\right)=0$, that leads to the following relation between the othoradial functions, $V_l$, and the radial one, $U_l$: $V_{l}\left(r\right)=\frac{1}{l\left(l+1\right)\rho r}\frac{{\rm d}\left(\rho r^2 U_{l}\right)}{{\rm d}r}$. Finally, $\vec u\left(r,\theta,\varphi,t\right)$ is the velocity field of the internal waves.\\

Next, the temperature, $T$, and the mean molecular weight, $\mu$, are respectively expanded as:
\begin{equation}
T\left(r,\theta\right)={\overline T}\left(r\right)+\delta T\left(r,\theta\right) \quad
\hbox{with} \quad \delta T\left(r,\theta\right)=\sum_{l\ge 2}\left[\Psi_{l}\left(r\right)\overline{T}\right]P_{l}\left(\cos\theta\right),
\end{equation}
\begin{equation}
\hbox{and}\hbox{ }\mu\left(r,\theta\right)={\overline \mu}\left(r\right)+\delta \mu\left(r,\theta\right) \quad
\hbox{with}\quad \delta \mu\left(r,\theta\right)=\sum_{l\ge 2}\left[\Lambda_{l}\left(r\right)\overline{\mu}\right]P_{l}\left(\cos\theta\right);
\end{equation}
${\overline T}$ and ${\overline \mu}$ are their horizontal averages, ${\delta T}$ and $\delta \mu$ being their fluctuations and $\Psi_l$ and $\Lambda_l$ their relative fluctuations.

Finally, the magnetic field is expanded using its divergence-free property:
\begin{equation}
\vec B\left(r,\theta\right)=\vec\nabla\wedge\vec\nabla\wedge\left(\xi_{P}\left(r,\theta\right){\widehat e}_{r}\right)+\vec\nabla\wedge\left(\xi_{T}\left(r,\theta\right){\widehat e}_{r}\right).
\end{equation}
$\xi_{P}$ and $\xi_{\rm T}$ are respectively the poloidal and the toroidal magnetic stream functions which are expanded in spherical harmonics:
\begin{equation}
\xi_{P}\left(r,\theta\right)=\sum_{l=1}^{\infty}\xi_{0}^{l}\left(r\right)Y_{l}^{0}\left(\theta\right)\hbox{ }
\hbox{and}\hbox{ }
\xi_{T}\left(r,\theta\right)=\sum_{l=1}^{\infty}\chi_{0}^{l}\left(r\right)Y_{l}^{0}\left(\theta\right).
\end{equation}
Here, the mean axisymetric magnetic field is considered ($m=0$). Therefore, the poloidal field, $\vec B_{P}\left(r,\theta\right)$, is in the meridional plane $\left({\widehat e}_{r},{\widehat e}_{\theta}\right)$ while the toroidal one, $\vec B_{T}\left(r,\theta\right)$, is purely azimuthal along ${\widehat e}_{\varphi}$.\\

The different fields being now defined, we have to consider the transport equations that have to be implemented in stellar evolution codes.

\subsection{Transport equations system}

The first equation of transport we consider is the one related to the transport of the magnetic field, namely the equation of induction:
\begin{equation}
\partial_{t}\vec B=\vec\nabla\wedge\left(\vec V\wedge\vec B\right)-\vec\nabla\wedge\left(||\eta||\otimes\vec\nabla\wedge\vec B\right),
\label{Ind}
\end{equation} 
$||\eta||$ being the magnetic eddy-diffusivity tensor. Two advection-diffusion equations, for respectively $\xi_{P}$ and $\xi_{T}$, are obtained from its spectral expansion in the vectorial spherical harmonics: 
\begin{eqnarray}
\frac{{\rm d}}{{\rm d}t}\xi_{0}^{l}\underbrace{-r{\mathcal P}_{{\rm \bf Ad};l}\left(U_{l},\vec B\right)}_{\rm Ia}=\underbrace{\eta_{h}r\Delta_{l}\left(\frac{\xi_{0}^{l}}{r}\right)}_{\rm IIa}\quad\quad\hbox{and}\nonumber
\end{eqnarray}
\begin{equation}
\frac{\rm d}{{\rm d}t}\chi_{0}^{l}+\partial_{r}\left(\dot r\right)\chi_{0}^{l}\underbrace{-{\mathcal T}_{{\rm \bf Ad};l}\left(\Omega,U_{l},\vec B\right)}_{\rm Ib}=\underbrace{\left[\partial_{r}\left(\eta_{h}\partial_{r}\chi_{0}^{l}\right)-\eta_{v}l\left(l+1\right)\frac{\chi_{0}^{l}}{r^2}\right]}_{\rm IIb}.
\end{equation}
The terms Ia and Ib, where ${\mathcal P}_{{\rm \bf Ad};l}\left(U_{l},\vec B\right)$ and ${\mathcal T}_{{\rm \bf Ad};l}\left(\Omega,U_{l},\vec B\right)$ are function of the differential rotation, the meridional circulation and the magnetic field, correspond to the advection of $\vec B$ by $\vec{\mathcal U}_{M}$ and to the production of its toroidal component  through the shear of the differential rotation. Terms IIa and IIb correspond to the turbulent ohmic diffusion, where the possibility of an anisotropic turbulent transport of the magnetic field is assumed, with $\eta_{v}$ and $\eta_{h}$ being respectively the eddy-magnetic diffusivity in the vertical direction and in the horizontal one. The transport is studied from a lagrangian point of view due to the contractions and to the dilatations of the star during its evolution. The time-lagrangian derivative ${\rm d}/{{\rm d}t}$ is defined by ${\rm d}/{\rm d}t=\partial_{t}+{\dot r}\partial_{r}$.\\

The second equation which has to be treated is the Navier-Stockes equation, in other words the equation of dynamics:
\begin{equation}
\rho\left[\partial_{t}\vec V+\left(\vec V\cdot\vec\nabla\right)\vec V\right]=-\vec\nabla P-\rho\vec\nabla\phi+\vec\nabla\cdot||\tau||+\left[\frac{1}{\mu_{0}}\left(\vec\nabla\wedge\vec B\right)\right]\wedge\vec B.
\label{Dyn}
\end{equation}
$P$, $\phi$ are respectively the pressure and the gravitational potential. $||\tau||$ is the Reynolds stress tensor and the last term is the Lorentz force where $\mu_{0}$ is the magnetic permeability of vaccum. Taking its azimuthal component and averaging over $\varphi$, the following advection-diffusion equation for the mean rotation rate on an isobar is obtained:
\begin{equation}
\rho\frac{{\rm d}}{{\rm d}t}\left(r^2\overline{\Omega}\right)\underbrace{-\frac{1}{5r^2}\partial_{r}\left(\rho r^4 \overline{\Omega}U_{2}\right)}_{\rm I}=
\underbrace{\frac{1}{r^2}\partial_{r}\left(\rho\nu_{v}r^4\partial_{r}\overline{\Omega}\right)}_{\rm II}
+\underbrace{\overline{\Gamma}_{{\vec{\mathcal F}}_{\mathcal L}}\left(\vec B\right)}_{\rm III}-\underbrace{\frac{1}{r^2}\partial_{r}\left[{\mathcal F}_{J}\left(r\right)\right]}_{\rm IV}\nonumber\\
\end{equation}
where $\overline{\Omega}\left(r\right)=\int_{0}^{\pi}\Omega\left(r,\theta\right)\sin^{2}\theta{\rm d}\theta/\int_{0}^{\pi}\sin^{3}\theta{\rm d}\theta$.
Term ${\rm I}$ corresponds to the advection of angular momentum by the meridional circulation and term ${\rm II}$ is the diffusive term associated to the action of the shear-induced turbulence where $\nu_v$ is the eddy-viscosity in the vertical direction. These two first terms correspond to the rotational transport of type I. Term ${\rm III}$ is associated to the Lorentz torque, $\overline{\Gamma}_{{\vec{\mathcal F}}_{\mathcal L}}$, and term ${\rm IV}$ corresponds to the transport by internal waves, ${\mathcal F}_{J}\left(r\right)$ being the associated mean angular momentum flux on an isobar. These two last terms correspond to the rotational transport of type II. The same type of equation is obtained for the differential rotation in latitude (cf. Mathis \& Zahn 2004-2005).

On the other hand, due to the long time-scales associated to the meridional circulation, dynamical terms in the meridional components of the Navier-Stockes equation are filtered, keeping only its hydrostatic terms. Taking the curl of the hydrostatic equation, we get the associated thermal-wind equation: 
\begin{equation}
\varphi\Lambda_{l}-\delta\Psi_{l}=\frac{r}{\overline g}{\mathcal D}_{l}\left(\Omega,\vec B\right),
\label{ThermalWind}
\end{equation}
where $\overline g$ is the horizontal average of gravity, the explicit form of ${\mathcal D}_{l}$ in function of $\Omega$ and $\vec B$ being given in Mathis \& Zahn 2005.
The more general equation of state is used (cf. Kippenhahn \& Weigert 1990) where $\delta=-\left(\partial\ln \rho/\partial\ln T\right)_{P,\mu}$ and $\varphi=\left(\partial\ln \rho/\partial\ln \mu\right)_{P,T}$. \\

The last equation that has to be solved is the one for the macroscopic entropy, $S$, in other words the equation for the transport of heat:
\begin{equation}
\rho T \left[\partial_{t}S+\vec V\cdot\vec\nabla S\right]=\vec\nabla\cdot\left(\chi\vec\nabla T\right)+\rho\epsilon-\vec\nabla\cdot\vec F+\mathcal J.
\label{Entropy}
\end{equation}
Its describes the transport of entropy by advection with taking into account the thermal diffusion ($\chi$ is the thermal conductivity), the production of energy associated to the nuclear reactions ($\epsilon$ is the nuclear energy production rate per unit mass), the heating due to turbulence $\vec\nabla\cdot\vec F=-\vec\nabla\cdot\left[\rho T ||D||\otimes\vec\nabla S\right]$ where $||D||$ is the eddy-diffusivity tensor and the ohmic heating ${\mathcal J}=\frac{1}{\mu_{0}}\left[||\eta||\otimes\left(\vec\nabla\wedge\vec B\right)\right]\cdot\left(\vec\nabla\wedge\vec B\right)$. Its expansion in spherical functions leads to the following equation for the transport of the temperature fluctuation:
\begin{equation}
C_{p}\overline{T}{{\rm d} \Psi_{l} \over {\rm d} t}+\underbrace{\Phi\frac{{\rm d}\ln\overline{\mu}}{{\rm d}t}\Lambda_{l}}_{\rm I}+\underbrace{\frac{U_{l}(r)}{H_{p}}\left(\nabla_{ad}-\nabla\right)}_{\rm II}=\frac{L\left(r\right)}{M\left(r\right)}{\mathcal T}_{l}(r)+\underbrace{\frac{\mathcal J_{l}}{\overline \rho}}_{\rm III},
\label{Mer}
\end{equation}
${\mathcal T}_{l}(r)$ being given by:
\begin{equation}
{\mathcal T}_{l}={\mathcal T}_{l;{\mathcal P}}\left(r\right)+{\mathcal T}_{l;{\rm Th}}\left(r\right)+{\mathcal T}_{l;{\rm N-G}}\left(r\right)
\end{equation}
where
\begin{eqnarray}
{\mathcal T}_{l;{\mathcal P}}&=&2\left[1-\frac{\overline{f}_{\mathcal{P}}\left(\Omega,\vec B \right)}{4\pi G\overline{\rho}}-\frac{\left(\overline{\epsilon}+\overline{\epsilon}_{\rm grav}\right)}{\epsilon_{m}}\right]\frac{\widetilde{g}_{l}\left(\Omega, \vec B \right)}{\overline{g}}+\frac{\widetilde{f}_{\mathcal{P},l}\left(\Omega, \vec B \right)}{4\pi G\overline{\rho}}\nonumber\\
&&-\frac{\overline{f}_{\mathcal{P}}\left(\Omega,\vec B\right)}{4\pi G\overline{\rho}}\left(-\delta\Psi_{l}+\varphi\Lambda_{l}\right),\nonumber
\end{eqnarray}
\begin{eqnarray}
{\mathcal T}_{l;{\rm Th}}=\frac{\rho_{m}}{\overline{\rho}}\left[\frac{r}{3}\partial_{r}X_{l}\left(r\right)
-\frac{l\left(l+1\right) H_{T}}{3 r}\left(1 + \frac{D_{h}}{K}\right){\Psi_l}\right]\nonumber
\end{eqnarray}
\begin{eqnarray}
\quad\hbox{and}\quad{\mathcal T}_{l;{\rm N-G}}=\frac{\left(\overline{\epsilon}+\overline{\epsilon}_{grav}\right)}{\epsilon_{m}}\left[X_{l}\left(r\right)+(f_{\epsilon}\epsilon_{T} - f_{\epsilon}\delta + \delta)\Psi_{l}+(f_{\epsilon}\epsilon_{\mu}+f_{\epsilon}\varphi - \varphi)\Lambda_{l}\right]\nonumber
\end{eqnarray}
with $X_{l}\left(r\right)=H_{T}\partial_{r}\Psi_{l}-(1-\delta+\chi_{T})\Psi_{l}-(\varphi+\chi_{\mu})\Lambda_{l}$. $L$ is the luminosity of the star, $M$ its mass and $K={\overline{\chi}}/{\overline{\rho}C_{p}}$ the thermal diffusivity. The definition of $\nabla$, $\nabla_{\rm ad}$, $C_{p}$, $H_{P}$, $H_{T}$, $f_{\epsilon}$, $\overline{\epsilon}$, $\overline{\epsilon}_{\rm grav}$, $\epsilon_{T}$, $\chi_{T}$, $\epsilon_{\mu}$, $\chi_{\mu}$, $\epsilon_{m}$ and $\rho_{m}$ are given in Mathis \& Zahn 2004-2005.Terms $\overline{f}_{\mathcal P}$, $\widetilde{f}_{\mathcal P,l}$ and $\widetilde{g}_{l}$, the fluctuation of the gravity on an isobar, are associated to the meridional perturbing force, $\vec{\mathcal F}_{\mathcal P}$, namely the sum of the centrifugal force, $\vec{\mathcal F}_{\mathcal C}=\frac{1}{2}\Omega^2\vec\nabla\left(r^2\sin^2\theta\right)$, and of the meridional Lorentz force, $\vec{\mathcal F}_{{\mathcal L},P}$. If we project $\vec{\mathcal F}_{\mathcal P}$ on the vectorial spherical harmonics, its explicit expansion in function of $\Omega$, $\xi_{l}$ and $\chi_{l}$ is derived (cf. Mathis \& Zahn 2005). The entropy of mixing is taken into account due to the multi-species caracteristic of the stellar plasma (cf. Maeder \& Zahn 1998), $\Phi$ being a function of the metal mass fraction $Z$ and of $\overline{\mu}$. Finally, ohmic heating has been expanded in spherical functions, ${\mathcal J}_{l}$ being the associated modal radial functions (their explicit expression in function of $\xi_{l}$ and $\chi_{l}$ is given in Mathis \& Zahn 2005).

Terms I, II and III in eq. \ref{Mer} correspond respectively to the multi-species characteristic of the stellar plasma, to the advection of entropy by the meridional circulation and to the ohmic heating.
Next, ${\mathcal T}_{l;{\mathcal P}}$ is the barotropic term of the meridional circulation which is generated by the perturbation of the thermal imbalance by $\vec{\mathcal F}_{\mathcal P}$. ${\mathcal T}_{l;{\rm Th}}$ corresponds to the thermal diffusion. In fact, if we keep its higher-order derivatives, we obtain $\frac{\rho_{m}}{\overline{\rho}}\frac{r}{3}\partial_{r}\left(H_{T}\partial_{r}\Psi_{l}\right)$ which is directly related with the temperature laplacian. Finally, ${\mathcal T}_{l;{\rm N-G}}$ is the term related to the nuclear reactions and to the heating due to the radial adjustements of the star during its evolution.\\

Finally, the equation for the transport of chemicals:
\begin{equation}
\rho\left[\partial_{t}c_i+\left(\vec V\cdot\vec\nabla\right)c_i\right]=\vec\nabla\cdot\left(\rho||D||\otimes\vec\nabla c_i\right)
\end{equation}
has to be solved to study the mixing; $c_i$ is the concentration of the i$^{\rm th}$ element which is considered. Expanding it on an isobar, it is obtained for the average of the concentration of each chemical, $\overline{c}_i$:
\begin{eqnarray}
\rho\frac{{\rm d}}{{\rm d}t}\overline{c}_{i}+\frac{1}{r^2}\partial_{r}\left[r^2\rho\overline{c}_{i}U_{i}^{\rm diff}\right]=\frac{1}{r^2}\partial_{r}\left[r^2\rho\left(D_{v}+D_{\rm eff}\right)\partial_{r}\overline{c}_{i}\right],
\label{Cmoy}
\end{eqnarray}
where $U_{i}^{\rm diff}$ is the velocity associated to the microscopic diffusion processes while the strong horizontal turbulence leads to the erosion of the advective transport which becomes a diffusive process (cf. Chaboyer \& Zahn 1992) with the following effective diffusion coefficient: $D_{\rm eff}=\sum_{l>0}\frac{\left(r U_l\right)^2}{l\left(l+1\right)\left(2l+1\right)D_{h}}$.\\

Next, taking the definition of the mean molecular weight:\\
$\frac{1}{\mu}=\sum_{i}\left[\left(1+Z_i\right)/A_i\right] c_i$ ($A_i$ and $Z_i$ are respectively the number of nucleons and of protons of the i$^{\rm th}$ element which is considered), the following advection-diffusion equation is obtained for $\Lambda_l$:
\begin{equation}
\frac{{\rm d}\Lambda_{l}}{{\rm d}t}-\frac{{\rm d}\ln \overline{\mu}}{{\rm d}t}\Lambda_{l}-\frac{U_l}{H_p}\nabla_{\mu}=-\frac{l\left(l+1\right)}{r^2}D_{h}\Lambda_{l}
\label{Lambda}
\end{equation}
where $\nabla_{\mu}=\frac{\partial \ln \overline{\mu}}{\partial \ln P}$.\\

Detailed boundary conditions associated to this complete set of partial differential equations is given in Mathis \& Zahn (2004) and in Mathis \& Zahn (2005).\\

These equations are all of advection-diffusion type. It has to be underlined that advection can not be treated as a diffusion process since it has been shown that solving the equation of dynamics begining with a flat rotation profile leads to a non-zero differential rotation, the advection being building angular velocity gradients (cf. Meynet \& Maeder 2000).\\

Moreover, note that making hydrostatic and thermal equilibrium hypothesis in \ref{Dyn} and \ref{Entropy} returns to the standard stellar evolution equations.

\section{Numerical simulation of secular transport: the hydrodynamical case with a 'shellular' rotation}

The numerical simulations presented in this work has been computed with the dynamical stellar evolution code STAREVOL and the reader is referred  to Siess et al. 2000, Palacios et al. 2003 and Palacios et al. 2006 for its detailed description. In the hydrodynamical case ($\vec B=\vec 0$) where we assume that the differential rotation is shellular, $\Omega\left(r,\theta\right)=\overline{\Omega}\left(r\right)$, due to the stronger horizontal turbulence which enforces the angular velocity to be constant on an isobar, the system of transport equations is reduced to:
\begin{equation}
\rho\frac{{\rm d}}{{\rm d}t}\left(r^2\overline{\Omega}\right)-\frac{1}{5r^2}\partial_{r}\left(\rho r^4 \overline{\Omega}U_{2}\right)=\frac{1}{r^2}\partial_{r}\left(\rho\nu_{v}r^4\partial_{r}\overline{\Omega}\right)
\end{equation}
and to the $l=2$ mode of Eqs. \ref{ThermalWind}, \ref{Mer} and \ref{Lambda} (here internal waves are not taken into account). It has been now implemented in STAREVOL. Associated boundary conditions which are used to compute simulations are detailed in Palacios et al. (2003).\\

\begin{figure}[h!]
\centering
\resizebox{13cm}{!}{\includegraphics{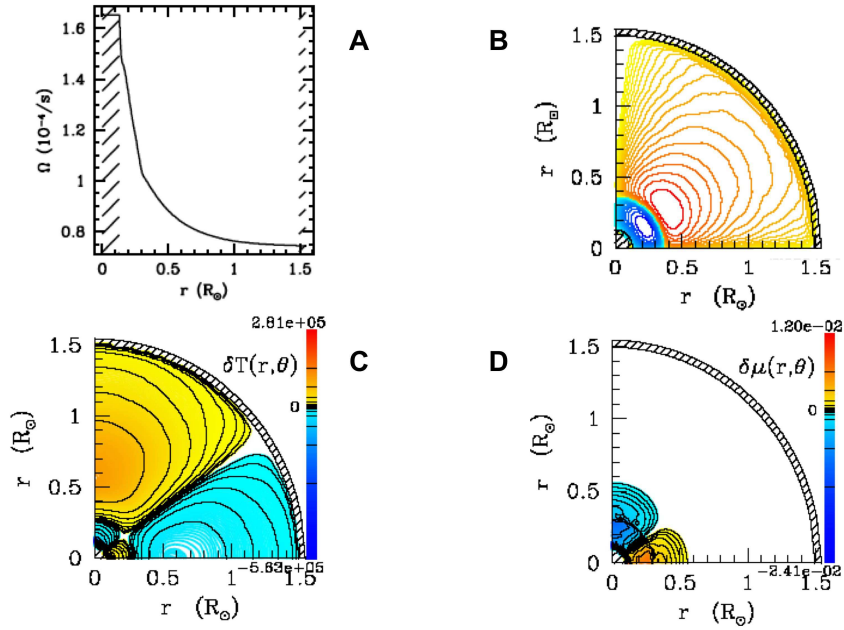}}
\caption{{\bf A:} The differential rotation profile. {\bf B:} Meridional circulation currents. In this model, the outer cell is turning counterclockwise allowing the equatorial extraction of angular momentum by the wind. {\bf C:} The T-excesses $\overline T\Psi_2 P_2\left(\cos\theta\right)$. It reachs $+2.81\times10^{5}$K in the inner region closer to the polar axis and $-5.63\times10^{5}$K in the inner equatorial region and become smaller near the surface. {\bf D:} The $\mu$-excesses $\overline\mu\Lambda_2 P_2\left(\cos\theta\right)$. They only occur close to the core and they are positive near the equatorial plane. Dashed lines regions correspond to the convective regions.}
\label{2D}
\end{figure}

The results presented here are issued from the numerical simulation of the evolution of a $1.5 M_{\odot}$ star with a solar metallicity $(Z=0.02)$ and an initial equatorial rotation velocity $V_{\rm ini}=100$ km.s$^{-1}$. The age of the presented model is $7.604 \times 10^{8}$ yr with a central Hydrogen mass fraction $X_c=0.57$. It is now possible to follow for each time-step the internal hydrodynamics of the radiation zone(s) of the star which is studied, following simultaneously the differential rotation profile (see Fig. \ref{2D}-A), the associated meridional circulation pattern (see Fig. \ref{2D}-B) and the temperature and the mean molecular weight fluctuations (see respectively Fig. \ref{2D}-C and \ref{2D}-D).\\

\begin{figure}[h!]
\centering
\resizebox{11.5cm}{!}{\includegraphics{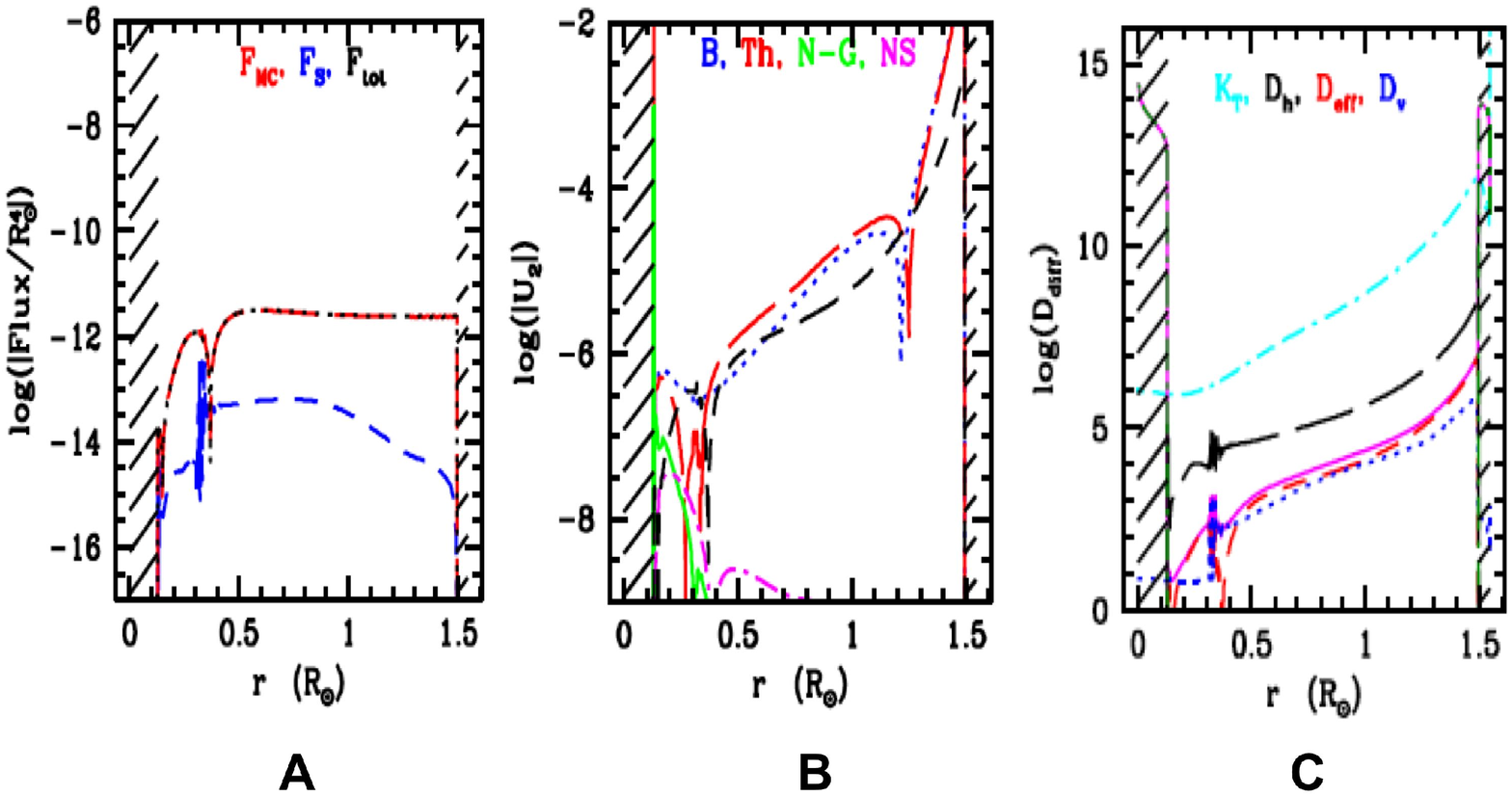}}
\caption{{\bf A:} Logarithm of the total flux of angular momentum (dots) and of that transported by meridional circulation, $F_{\rm MC}\left(r\right)=\frac{1}{R_{\odot}^{4}}\frac{1}{5}\rho r^4\overline{\Omega}U_{2}$, (continuous line) and by shear-induced turbulence, $F_{\rm S}\left(r\right)=\frac{1}{R_{\odot}^{4}}\rho\nu_{v}r^4\partial_{r}\overline{\Omega}$, (dashed line). {\bf B:} Logarithm of the meridional circulation (dashed line), the barotropic term (dots), the thermal diffusion term (long dashed line), the nuclear energy production and heating due to gravitational adjustements term (continuous line) and the non-stationarity term (dashed dotted line) profiles. {\bf C:} Logarithm of the themal diffusivity (dashed dotted line), the horizontal eddy-diffusivity (long dashed line), the effective diffusivity associated to meridional circulation (dashed line), the vertical eddy-diffusivity (dots) and the total vertical diffusivity, $D_{t}=D_{v}+D_{\rm eff}$, (continuous line) profiles. Dashed lines regions correspond to the convective regions.}
\label{FigDiag}
\end{figure}

Moreover, diagnosis tools have been developed to identify what are the dominant processes in the angular momentum transport, the meridional circulation and the chemicals mixing.\\
Considering Fig. \ref{FigDiag}-A, one can easily identify that here angular momentum transport is dominated by the advection by the meridional circulation, its flux transported by the shear-induced turbulence being smaller at least by an order of magnitude except in the region near the center. Then, looking at  Fig.\ref{FigDiag}-B, it can be easily identified that meridional circulation is mainly driven by the barotropic and the thermal diffusion terms. The term due to the nuclear energy production has to be taken into account only in the region of the star where nuclear reactions occur, here in the center, while the non-stationary term is completely negligible since the star which is studied here is a main-sequence star where structural adjustements are weak. Finally, concerning the transport coefficients, it can be seen that the meridional circulation is the dominating process in the transport of chemicals while our fundamental hypothesis concerning the turbulent transport is verified ($D_v\!<\!\!<\!D_h$) (cf. Fig. \ref{FigDiag}-C).\\
\newpage
Work is now in progress to implement the differential rotation in latitude and the transport by the magnetic field and the gravito-inertial waves, those two last processes being crucial to understand the internal transport of angular momentum in the Sun and the properties of low-masses stars.

\section{Asteroseismic diagnosis}
\begin{figure}[h!]
\centering
\resizebox{10cm}{!}{\includegraphics{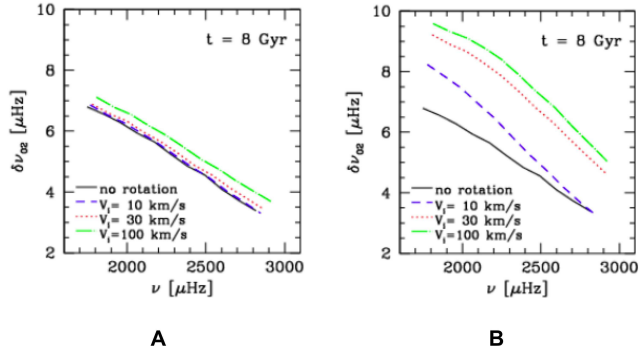}}
\caption{Value of the small difference $\delta_{0,2}$ in function of the frequency $\nu$ for different values of the initial rotation velocity, ${\rm V}_{\rm  i}$ for a $1{\rm M}_{\odot}$ star. {\bf A:} Case of the prescription of $\nu_{h}$ of Zahn (1992). {\bf B:} Case of the prescription for $\nu_{h}$ of Maeder (2003). (From Eggenberger Ph. D. Thesis)}
\label{AsteroDiag}
\end{figure}

Since theoretical results which contain some parametrization related to the internal dynamical processes are implemented in a stellar evolution code, constraints have to be obtained from observations on stellar interiors. Asteroseismology is the best way to get such constraints. In fact, rotational transports affects the stellar structure during the evolution of the star and thus oscillations frequencies are modified. If we recall the classical asymptotic relation for the frequencies of p-modes:
\begin{equation}
\nu_{n,l}\approx\left(n+l/2+\varepsilon\right)\Delta\nu-\frac{l\left(l+1\right)}{6}\delta\nu_{0,2},
\end{equation}
where $\Delta\nu$ and $\delta_{l,l+2}$ are respectively the large and the small separation:
\begin{eqnarray}
\Delta\nu=\left(2\int_{0}^{R}\frac{{\rm d}r}{c_s}\right)^{-1}\quad\hbox{and}\quad\delta_{l,l+2}=-\left(4l+6\right)\frac{\Delta\nu}{4\pi^2\nu_{n,l}}\int_{0}^{R}\frac{{\rm d}c_{s}}{{\rm d}r}\frac{{\rm d}r}{r},\nonumber
\end{eqnarray}
differences are obtained if we consider a non-rotating or a rotating model. Moreover, constraints could be obtained for each specific transport process by studying the effect of its different modelling which could be done on seismic observables. For example, if models are computed using the different prescriptions which exist for the horizontal eddy-viscosity, $\nu_{h}$, the effect on the small separation once the star has evolved is important with an increase associated with that of the mean rotation rate of the star and of $\nu_{h}$ which leads to a more important transport and thus to a stronger rotational transport's indirect effect on frequencies. Here, results are reported for a $1 M_{\odot}$ star with different initial rotation velocities , $V_i=0,10,30,100\hbox{ }{\rm km.s^{-1}}$ at $t=8$ Gyr (cf. Fig. \ref{AsteroDiag}). The models are built using the solar calibration and the different prescriptions for the horizontal turbulence established by Zahn (1992), Maeder (2003) and Mathis et al. (2004). Simulations are performed using the Geneva stellar evolution code (cf. Meynet \& Maeder 2000) where the microscopic diffusion has been implemented in the same way that in Richard et al. (1996), the braking being treated using the formalism of Kawaler (1988). The asteroseismic analysis is performed using the Aarhus oscillation code (cf. Christensen-Dalsgaard 1997).

\section{Conclusion}
In this work, a coherent description of the dynamical transport processes which take place in stellar radiation zones has been undertaken. Each of them and their respective effects on angular momentum and chemical transport has been identified and modelled in a consistent way. For the first time, a two-dimensional picture of the internal dynamics of stellar radiation zones is obtained and the first step of the numerical implementation of the theoretical results in stellar evolution codes, namely the purely hydrodynamical case where it is assumed that the strong horizontal turbulent transport enforces a shellular rotation law, has been achieved. Moreover, work is in progress to implement the differential rotation in latitude, the magnetic field, the gravito-inertial waves and the tides while theoretical work is engaged to derive prescriptions for the MHD instabilities, the waves excitation and the tides dissipation. On the other hand, we have shown how classical seismic observables are affected by the internal dynamical processes that will allow to put constraints on our modelling from the comparison with the observations. Thus, with the first light of COROT, the preparation of KEPLER, SOHO and forthcoming helioseismic space instrument such as GOLF-NG on DynaMICCS project (Turck-Chi\`eze et al. 2005, Turck-Chi\`eze et al. 2006), we are now entering in a new exciting period of the story of stellar evolution where we hope to be able to obtain a dynamical vision of the Hertzsprung-Russel diagram with the support of helio and asteroseismology.

%%-----------------------------
%%      your bibliography
%%-----------------------------

\end{document}